\documentclass[pre,aps,twocolumn,superscriptaddress]{revtex4}
\usepackage{amsmath, amsfonts, amssymb, graphicx, color}

\newcommand{\TM}{}

\newcommand{\GM}{}
\newcommand{\<}{\langle}
\renewcommand{\>}{\rangle}

\newcommand{\UF}[1]{{#1}}

\begin{document}

 % title
\newcommand{\deftitle}{Strong, but not weak, noise correlations are beneficial for population coding}

\title{\deftitle}

\author{Gabriel Mahuas}
\affiliation{Institut de la Vision, Sorbonne Université, CNRS, INSERM, 17 rue Moreau, 75012, Paris, France}
\affiliation{Laboratoire de physique de \'Ecole normale sup\'erieure,
  CNRS, PSL University, Sorbonne University, Universit\'e Paris-Cit\'e, 24 rue Lhomond,
  75005 Paris, France}
\author{Thomas Buffet}
\affiliation{Institut de la Vision, Sorbonne Université, CNRS, INSERM, 17 rue Moreau, 75012, Paris, France}
\author{Olivier Marre}
\affiliation{Institut de la Vision, Sorbonne Université, CNRS, INSERM, 17 rue Moreau, 75012, Paris, France}
\author{Ulisse Ferrari}
\thanks{These authors contributed equally.}
\affiliation{Institut de la Vision, Sorbonne Université, CNRS, INSERM, 17 rue Moreau, 75012, Paris, France}
\author{Thierry Mora}
\thanks{These authors contributed equally.}
\affiliation{Laboratoire de physique de \'Ecole normale sup\'erieure,
  CNRS, PSL University, Sorbonne University, Universit\'e Paris-Cit\'e, 24 rue Lhomond,
  75005 Paris, France}

\begin{abstract}
 % abstract
Neural correlations play a critical role in sensory information coding. They are of two kinds: signal correlations, when neurons have overlapping sensitivities, and noise correlations from network effects and shared noise. 
\UF{In experiments from early sensory systems and cortex, many pairs of neurons typically show both types of correlations to be positive and large, especially between nearby neurons with similar stimulus sensitivity. 
However, theoretical arguments have suggested that stimulus and noise correlations should have opposite signs to improve coding, at odds with experimental observations.
We analyze retinal recording in response to a large variety of stimuli, and show that, contrary to common belief, large noise correlations are beneficial for coding, even if aligned with signal correlations.
To understand this result, we develop a theory of visual information coding by correlated neurons, which resolves that paradox.
We show that noise correlations are always beneficial if they are strong enough, unless neurons are perfectly correlated by the stimulus.}
Finally, using neuronal recordings and modeling, we show that for high dimensional stimuli noise correlation benefits the encoding of fine-grained details of visual stimuli, at the expense of large-scale features, which are already well encoded.

\end{abstract}

\maketitle

\section{Introduction}

 % intro
 Neurons from sensory systems encode information about incoming stimuli in their collective spiking activity.
This activity is noisy: repetitions of the very same stimulus can drive different responses \cite{Zohary94,Meister95,Gawne93,Smith08,Hofer11,Cohen11}.
It has been shown that the noise is shared among neurons and synchronizes them, an effect called {\em noise correlations}, as opposed to {\em signal correlations} induced by the stimulus \cite{Averbeck2006review,Cohen11,Kohn16,Azeredo21,Panzeri22}.
Noise correlations have been observed since the first synchronous recordings of multiple neurons \cite{Perkel67,Mastronarde89} and at all levels of sensory processing, from the retina \cite{Meister95,Nirenberg01,Shlens08,Pillow08,Volgyi09,Volgyi13,Franke16,Zylberberg16,Ruda20,Sorochynskyi21} to the visual cortex \cite{Zohary94,Fiser04,Kohn05,Smith08,Ecker10,Hofer11,Lin15} and other brain areas \cite{Usrey99,Averbeck2006review,Lee98,Bair01,Cohen11,Azeredo21,Hazon22}

Strong noise correlations have been measured mostly between nearby neurons with similar {stimulus sensitivity} \cite{Mastronarde83,Mastronarde89,Zohary94,Lee98,Bair01,Kohn05,Gutnisky08,Averbeck06,Hofer11}. {This behaviour is }particularly evident in the retina between nearby ganglion cells of the same type \cite{Meister95,Shlens08,Volgyi09,Volgyi13,Sorochynskyi21}.
This observation {is however surprising, since previously it was thought that these correlations are detrimental to information coding: a theoretical argument \cite{Zohary94,Brunel98,Panzeri99,Pola03,Sompolinsky01} }suggests that noise correlations are detrimental to information transmission if they have the same sign as signal correlations \cite{Averbeck2006review,Azeredo21,Panzeri22}. 
This rule is sometimes called the sign rule \cite{Hu14}, and is related to the notion of information-limiting correlations \cite{Moreno-Bote14}.
Since nearby neurons with similar tuning are positively correlated by the signal, the theory would predict that their positive noise correlations should be detrimental, making the code less efficient. However, a large body of literature has reported the beneficial effects of noise correlations on coding accuracy{\TM, with various approaches and reasonings: by computing the Fisher information \cite{Abbott99,Ecker11} or mutual information \cite{Tkacik10,Nardin23} in models of large populations of correlated neurons;
by studying specific cases where the sign rule applies \cite{Azeredo14,Franke16}; by showing explicitly that decorrelating noise reduces decoding accuracy \cite{Boffi2024};
or by showing that ignoring noise correlations in the decoder is detrimental \cite{Graf11,Pillow08,Ruda20}.}
Because of these contradictions, the effect of shared variability on information transmission is still unclear, and remains a largely debated topic in neuroscience \cite{Averbeck2006review,Kohn16,Azeredo21}.

Here we aim to resolve these tensions by developing a general framework that \UF{is grounded in the analysis of multi-electrode array recordings of rat and mouse retinas, and builds on previous theoretical work \cite{Mahuas23}.}
While previous studies have considered the impact of noise correlations either for particular stimuli \cite{Zohary94,Pillow08,Franke16,Hazon22}, or for particular models \cite{Sompolinsky01,Tkacik10,Ecker11}, our approach is general and covers both low and high dimensional stimuli. 
{We show that the sign rule can be broken in a specific regime that we observed in retinal responses: when noise correlations are strong enough compared to signal correlations, they have a beneficial effect on information transmission.
Our results unravel the complex interplay between signal and noise correlations, and predict when and how noise correlations are beneficial or detrimental.
In the case of high dimensional stimuli, like images or videos, our theory predicts different effects of noise correlations depending on stimulus features. 
In particular, it explains how large noise correlations between neurons with similar stimulus sensitivity help encode fine details of the stimulus.}

\UF{We first analyse the impact of correlated activity of rat retinal ganglion cells for a variety of stimuli and conditions. We observe violations of the sign rule, with large noise correlations being beneficial even for positive signal correlations.
We then build on \cite{Mahuas23} to develop a theory of information coding by correlated neurons which extends previous theoretical efforts and accounts for the observed regimes beyond the sign rule.} We extend our analysis to large populations of sensory neurons, and propose a spectral analysis suggesting that local noise correlations {enhance information by favoring} the accurate encoding of fine-grained details. Finally we validate that prediction by combining data from the mouse retina with accurate convolutional neural network (CNN) models.

\section{Results}

 % pairwise
\subsection*{Benefit of noise correlations in pairs of retinal ganglion cells}

 % fig1
\begin{figure*}
\begin{center}
\includegraphics[width=\textwidth]{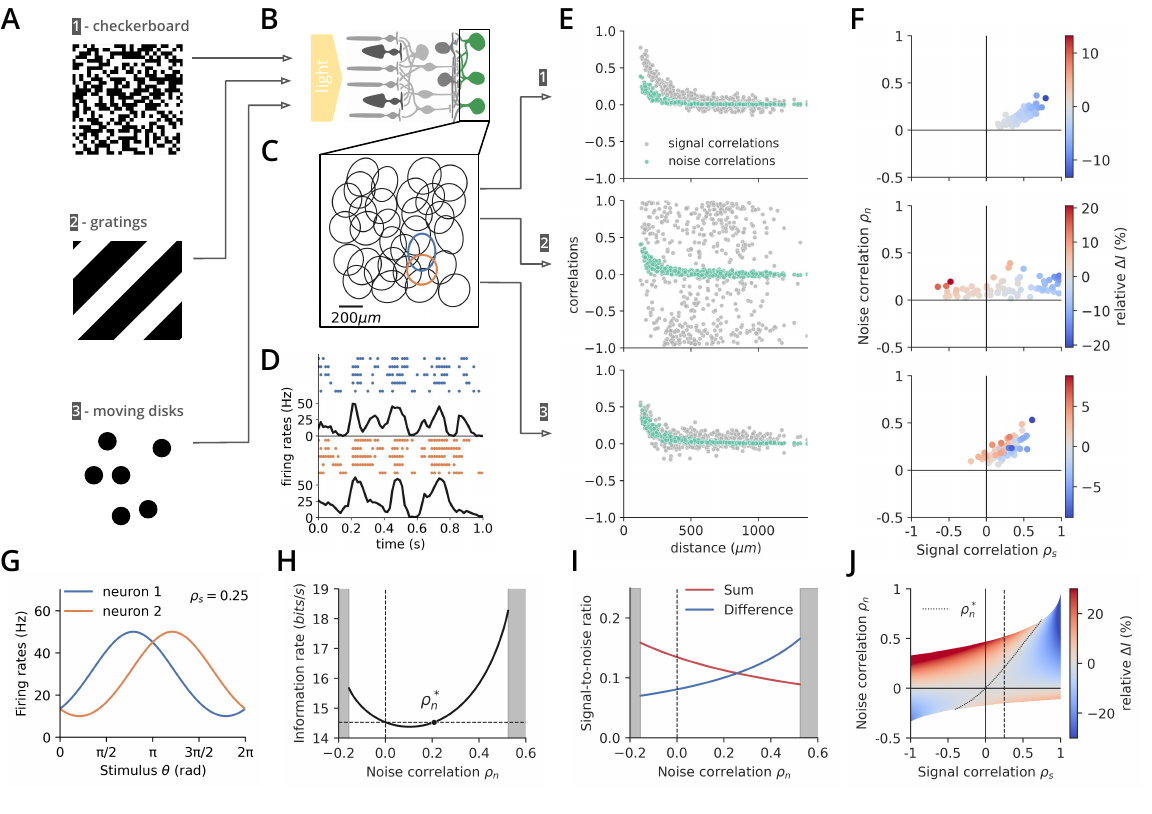}
\caption{{\bf The effects of noise correlations on information coding depends on the stimulus}. 
{\bf A.} Three stimuli with different spatiotemporal statistics were presented to a rat retina. 
{\bf B.} Retinal ganglion cells (RGCs) were recorded using a multi-electrode array (MEA). 
{\bf C.} We isolated a nearly complete population of OFF-$\alpha$ cells, with receptive fields (RFs) that tile the visual field following approximately a triangular lattice. 
{\bf D.} Example raster plots and firing rates for two cells with neighboring RFs. 
{\bf E.} Signal and noise correlations for each pairs of neurons in the population, versus their distance. Each plot corresponds to 1 of the 3 stimuli of E. 
{\bf F.} Noise synergy induced by noise correlations for all pairs of nearby neurons ($\geq 300 \mu m$), for each stimulus of E.
{\bf G.} Example pair of Von Mises tuning curves with moderate signal correlation level ($\rho_s=0.25$). 
{\bf H.} Mutual information between stimulus and response for the example pair of A, vs the strength of noise correlations. Grey areas correspond to forbidden correlations zones. 
{\bf I.} {The non-monoticity of B may be explained by examining how well the stimulus is represented by the sum and difference of the two neurons' activities, as measured by their signal-to-noise ratios. Noise correlations enhance noise in the sum, but reduce it in the difference.}
{\bf J.} Heatmap representing the noise synergy, defined as the relative gain of mutual information induced by noise correlations compared to the uncorrelated case. The dotted vertical line corresponds to the example pair of A and B.}
\label{fig:fig1}
\end{center}
\end{figure*}

{We first asked under what conditions noise correlations could be beneficial or detrimental in a population of sensory neurons, taking the retina as a test case}. We recorded {\em ex vivo} the joint spiking activity of rat retinal ganglion cells (RGCs, see Methods) \cite{Deny17, Botella18}. 
We subjected the same retinal preparation to 3 stimuli with distinct spatio-temporal patterns: a random flickering checkerboard, drifting gratings, and randomly moving disks (Fig.~\ref{fig:fig1}A). 
The activity of RGCs was recorded using a multi-electrode array (Fig.~\ref{fig:fig1}B), and data was processed to assign spikes to each neuron \cite{Yger18}. {\TM The time-dependent response was then defined as the number of spikes in each 20-ms time windows (chosen to capture the peak of noise correlations in pairwise cross-correlograms, see Supplemental Material Fig. S3).}
We identified cells belonging to a nearly complete OFF-$\alpha$ population forming a regular mosaic pattern of their receptive fields (Fig.~\ref{fig:fig1}C).

Each of the 3 stimulus movies was repeated multiple times (Fig.~\ref{fig:fig1}D),
which allowed us to compute the noise and signal correlation functions $\rho_{\rm n}$ and $\rho_{\rm s}$ (Fig.~\ref{fig:fig1}E), see Methods.
All three stimuli produced similar structures of noise correlations across the network, with positive correlations between cells with nearby receptive fields. This is consistent with the fact that noise correlations are a property of the network, independent of the stimulus \cite{Sorochynskyi21,Mahuas20}, and likely come here from gap junctions coupling neighbouring RGCs \cite{Brivanlou98,Volgyi09}. In contrast, signal correlations strongly depend on the statistical structure of the presented stimulus, and may be positive or negative, with varying strengths.

{We call ``noise synergy'' the gain in information afforded by noise correlations, $\Delta I=I(\rho_{\rm n})-I(\rho_{\rm n}=0)$.
To estimate it, we} compute {exactly} the mutual information between stimulus and response for all pairs of cells whose receptive fields were closer than $300\,\mu{\rm m}$, {and then subtract the same quantity computed on the data after shuffling across repetitions} (Fig.~\ref{fig:fig1}F). 
The case of the drifting gratings {with fixed orientation} offers an illustration of the sign rule. That stimulus induces strong {positive or} negative signal correlations between many cells, {depending on their relative positions relative to the gratings direction}. Since noise correlations are positive, they may be either of the same sign as signal correlations, and therefore detrimental, or of opposite sign, hence beneficial. 
In the case of the checkerboard stimulus, noise correlations were found to be generally detrimental. This again agrees with the sign rule since they have the same sign as signal correlations. 
Finally, the case of the moving disks provides an example of {a} third regime, which violates the sign rule:
noise correlations are of the same sign as the signal correlations, but also of comparable magnitude; yet they are beneficial for information. {\TM We checked that those results were robust to the choice of bin (Supplemental Material Fig. S3)}.

Overall, the 3 stimuli illustrate 3 possible regimes when noise correlations are positive: a beneficial effect when signal correlations are negative, a detrimental effect if signal correlations are positive and {large}, and a beneficial effect when noise and signal correlations are both positive and of the same magnitude.

\subsection*{Strong pairwise noise correlations enhance information transmission}

{To make sense of our experimental findings, we} consider a simple model of a pair of spiking neurons encoding an angle $\theta$, for instance the direction of motion of a visual stimulus, in their responses $r_1$ and $r_2$.
These responses are correlated through two sources: signal correlations $\rho_{\rm s}$ due to an overlap of the tuning curves (Fig. \ref{fig:fig1}G); and {constant} noise correlations $\rho_{\rm n}$ due to shared noise (see Methods for mathematical definitions).
We asked how this shared noise affects the encoded information, {for a fixed level of noise in neurons}.

To quantify the joint coding capacity of the 2 neurons, we computed {exactly} the mutual information $I(\theta;r_1,r_2)$ between their activities and the stimulus $\theta$.
For fixed tuning curves, we find that the mutual information depends non monotonously on the noise correlation $\rho_{\rm n}$ (Fig. \ref{fig:fig1}H). 
For small absolute values of $\rho_{\rm n}$, the sign rule is satisfied, meaning that negative noise correlations are beneficial, and weak positive ones are detrimental \cite{Hu14, Averbeck2006review, Abbott99, Pola03}.
However, the mutual information increases again and noise correlations become beneficial if they are larger than a certain threshold $\rho_{\rm n}^*$, violating the sign rule.
{This non monotonous dependency may be intuitively explained as the interplay between two opposite effects (Fig.~\ref{fig:fig1}I). Negative noise correlations are beneficial because they reduce noise in the total activity of the neurons. By contrast, positive noise correlations reduce noise in their differential activity, but this effect only dominates when they are strong enough.}

{Mirorring the experimental results of Fig.~\ref{fig:fig1}F,} Fig.~\ref{fig:fig1}J shows theoretically how noise synergy depends on both the noise and signal correlation, where the latter is varied in the model by changing the overlap between the tuning curves. 
Very generally, and beyond the cases predicted by the sign-rule, noise correlations are beneficial also when they are stronger than the signal correlations, {as we had observed in the retinal data (bottom plot of Fig.~\ref{fig:fig1}F).}
We can gain insight into this behaviour by computing an approximation of the mutual information that is valid for small correlations, following \cite{Mahuas23} (see Methods).
The noise synergy can be expressed as:
\begin{equation}\label{eq:dMI_2cd}
  \Delta I\approx\frac{\alpha}2 \rho_{\rm n}(\rho_{\rm n}-\rho_{\rm n}^*),
\end{equation}
where $\alpha \leq 1$ is prefactor that grows with the signal-to-noise ratio (SNR) of the neurons.
Eq.~\ref{eq:dMI_2cd} captures the behaviour of Fig.~\ref{fig:fig1}{F and J}, in particular the observation that noise correlations are beneficial if $\rho_{\rm n}\rho_{\rm s}<0$, as the sign rule predicts, or if they are strong enough, $|\rho_{\rm n}|>|\rho_{\rm n}^*|$. We can show (see Methods) that the threshold $\rho_{\rm n}^*$ scales with the signal correlation strength $\rho_{\rm s}$:
\begin{equation}\label{eq:rhostar}
\rho_{\rm n}^* = \beta\, \rho_{\rm s}.
\end{equation}
{This result holds for either discretely spiking or Gaussian neurons (see Methods) and} the prefactor $\beta\leq 1$ gets smaller and even approaches 0 as the SNR increases. It is also smaller when these SNR are dissimilar between cells, consistent with previous reports~\cite{Ecker11}. When the SNRs are weak and similar, we have $\beta\approx 1$. This analysis indicates that noise correlations are beneficial when they are of the same strength as signal correlations, but also that this benefit is enhanced when neurons are reliable.

{Our definition of the noise synergy relies on comparing the noise-correlated and uncorrelated cases at fixed noise level or SNR.
However, increasing noise correlations at constant SNR decreases the effective variability of the response, as measured by the noise entropy of the joint response of the pair (see Methods). This means that high noise correlations imply a more precise response, which could explain the gain in information. To study this possible confounding factor, we also computed $\Delta I$ at equal noise entropy, instead of equal SNR, and found that strong noise correlations are still beneficial, with modified $\rho^* = 2 \rho_{\rm s}/(1+\rho_{\rm s}^2)\leq 1$ (see Methods).}

Our theoretical predictions rely on the assumption that noise (Pearson) correlations do not vary with the stimulus. We wondered whether the theory was robust to stimulus-dependent noise correlations. Previous work based on Fisher, rather than Shannon, information suggested that noise correlations are detrimental when aligned to the signal direction in each point of response space \cite{Moreno-Bote14, Kanitscheider15}.
This structure, called ``differential'' or ``information-limiting'' correlations, implies that noise correlations vary as a function of the stimulus itself. Their detrimental effect can be intuited from the definition of the Fisher information \cite{Moreno-Bote14}, and in fact provide a sort of worst case scenario for the effect of noise correlations, as the alignment of noise and signal is satisfied locally for all values of the stimulus, and thus maximally detrimental according to the sign rule.
To investigate this scenario, we performed a numerical analysis of the noise synergy in a two-neuron system with information-limiting correlations (Supplemental Material, Appendix A and Fig.~S1A).
We find that information-limiting correlations become increasingly beneficial to the mutual information as their strength increases (Supplemental Material, Fig.~S1B), while they are always detrimental to the Fisher information (Supplemental Material, Fig.~S1C), consistent with previous results. While perhaps counterintuitive, this result can be understood by considering the effect of the curvature of the response curve, which induces non-local, non-linear effects (ignored by the Fisher information) that become stronger with increasing noise correlations (Supplemental Material, Fig.~S1A).
Again, it's worth stressing that information-limiting correlations were designed to be maximally detrimental. Our finding that even those are benefical as long as noise correlation are strong enough provide strong evidence that our conclusion is very general, the only exception being the trivial case of perfectly correlated neurons ($\rho_s=1$).

 % pop
\subsection*{Large sensory populations {in high dimension}}

 % fig2
\begin{figure*}
\begin{center}
  \includegraphics[width=1.\textwidth]{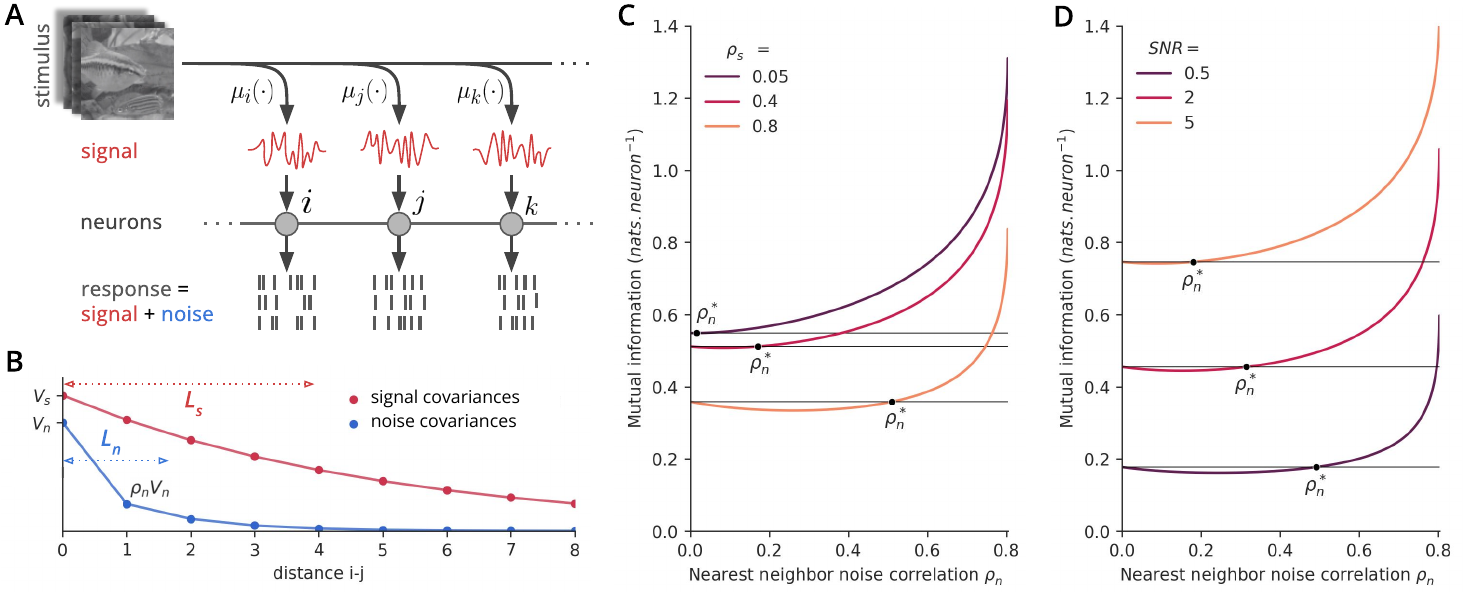}
\end{center}
\caption{
  {\bf Population analysis}. {\bf A.} Neurons are assumed to be spatially arranged along sensory space. They combine features of the stimulus through a response function $\mu_i$. Noise is added to the neural responses. {\bf B.} Signal and noise covariances versus distance between neurons. Signal and noise covariances decay exponentially with distance with spatial scales $L_{\rm{s}}$ and $L_{\rm{n}}$. {\bf C.-D}. Mutual information as a function of the noise correlation between neighbors for: ({\bf C}) varying levels of signal correlations, with fixed $V_{\rm{s}}=2$, $V_{\rm{n}}=1$, and $L_{\rm{n}}=2$; and ({\bf D}) varying levels of signal-to-noise ratio (SNR$=V_{\rm{s}}/V_{\rm{n}}$), with $L_{\rm{s}}=L_{\rm{n}}=2$ ($\rho_{\rm{s}}\approx 0.6$).
}
\label{fig:fig2}
\end{figure*}

{We then asked how these results extend from pairs to large populations,
by considering a large number of neurons tiling sensory space (Fig.~\ref{fig:fig2}A).}
{To go beyond neurons tuned to a single stimulus dimension, and account for the ability of neurons to respond to different stimuli in a variety of natural contexts, we assume that each neuron responds to high-dimensional stimulus, like a whole image, a temporal sequence, or a movie.}
As different stimuli are shown, the spike rate of each neuron will vary. For computational ease, we take these fluctuations to be Gaussian of variance $V_{\rm s}$.

{To account for the empirical observation that nearby neurons tend to have close receptive fields, we correlate the responses of any two neurons with a strength that decreases as a function of their distance in sensory space, with characteristic decay length $L_{\rm s}$ (Fig.~\ref{fig:fig2}B).}
The value of the correlation between nearest neighbours quantifies the signal correlation, $\rho_{\rm s}$.
For simplicity the response noise is also assumed to be Gaussian of variance $V_{\rm n}$.
{To model positive noise correlations between nearby neurons observed in both the retina \cite{Meister95,Shlens08,Volgyi09,Volgyi13,Sorochynskyi21} and cortex \cite{Mastronarde83,Mastronarde89,Zohary94,Lee98,Bair01,Kohn05,Gutnisky08,Averbeck06,Hofer11}, we assume that they also decay with distance, but with a different length $L_{\rm n}$ (Fig.~\ref{fig:fig2}B). 
The noise correlation between nearest neighbors, defined as $\rho_{\rm n}$, quantifies their strength.}

In this setting, both signal and noise correlations are positive, and the sign rule alone would predict a detrimental effect of noise correlations.
The mutual information can be computed analytically in terms of simple linear algebra operations over the neurons' covariance matrices (see Methods) \cite{Atick90}.
Using these exact formulas, we examined how the mutual information changes as a function of the noise correlation $\rho_{\rm n}$ for different values of the signal correlation $\rho_{\rm s}$ (Fig.~\ref{fig:fig2}C) and of the SNR $V_{\rm s}/V_{\rm n}$ (Fig.~\ref{fig:fig2}D).

The results qualitatively agree with the case of pairs of neurons considered previously. 
Weak noise correlations impede information transmission, in accordance with the sign rule. 
However, they become beneficial as they increase past a critical threshold ($\rho_{\rm n}^*$), and this threshold grows with the signal correlation strength. It also decreases and even vanishes as the SNR is increased (Fig.~\ref{fig:fig2}D and Methods for a discussion of the large SNR limit). This means that more reliable neurons imply an enhanced benefit of noise correlations. We further proved that, even at low SNR, there always exists a range of noise correlation strengths where noise correlations are beneficial (see Methods). The general dependency of $\rho_{\rm n}^*$ on the correlation ranges $L_{\rm s}$ and $L_{\rm n}$ is shown in Supplemental Material Fig.~S2.

Based on the analysis of pairs of neurons, we expect inhomogeneities in the SNR $V_{\rm s}/V_{\rm n}$ of neurons to enhance the benefit of noise correlations. To study this effect, we let the power of the signal $V_{\rm s}$ vary between cells, while the noise level $V_{\rm n}$ is kept constant. 
Assuming that each cell is assigned a random value of $V_{\rm s}$, we can compute the correction to the critical noise correlation $\rho_{\rm n}^*$.
We find that $\rho_{\rm n}^*$ decreases at leading order with the magnitude of the inhomogeneity (see Methods). 
This result confirms that, in large populations of neurons as well, variability among neurons makes it more likely for noise correlations to have a beneficial effect.

 % spectral
\subsection*{Spectral decomposition} 

 % fig3
\begin{figure*}
\begin{center}
\includegraphics[width=1.\textwidth]{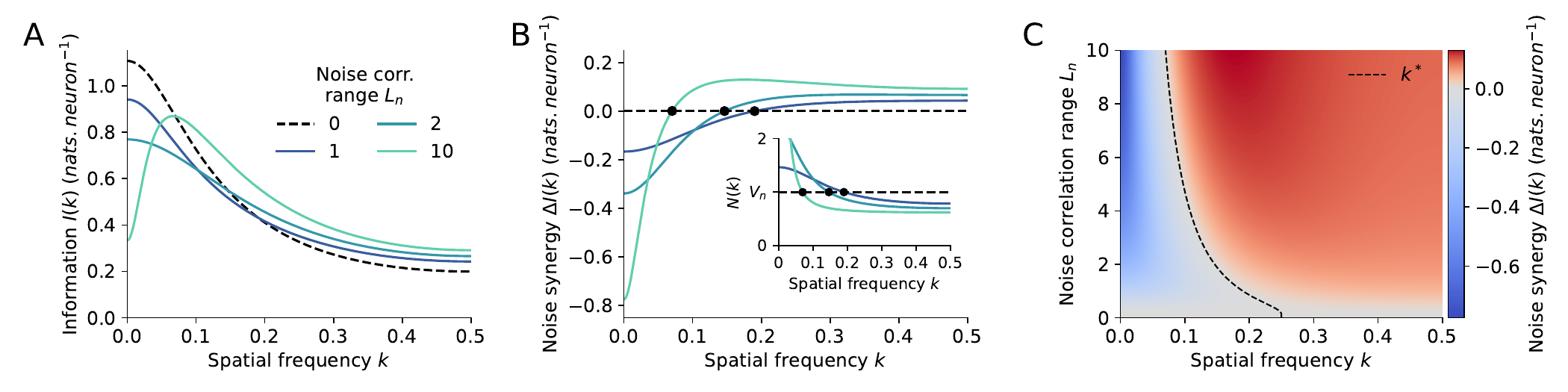}
\end{center}
\caption{
{\bf Spectral analysis of stimulus information encoding}. {\bf A.} Spatial spectral decomposition $I(k)$ of the mutual information between stimulus and response for a system with SNR$=2$, $L_{\rm{s}}=2$ and $\rho_{\rm{n}}=0.4$, for various ranges of the noise correlations ($L_{\rm{n}}=0$ corresponds to the absence of noise correlations). {\bf B.} Spectral decomposition of the noise synergy $\Delta I(k)=\log[({1+S(k)/N(k)})/({1+S(k)/V_{\rm n}})]$. The inset shows the power spectrum of the noise. {\bf C.} Heatmap showing the noise synergy spectral decomposition as a function of the noise correlation range $L_{\rm{n}}$. The critical spatial frequency $k^*$ above which noise correlations are beneficial is shown as a black dotted line.
}
\label{fig:fig3}
\end{figure*}

Mutual information is a single number that provides a global quantification of coding efficiency, but says nothing about what is being transmitted. 
Likewise, a positive noise synergy indicates that noise correlations are beneficial overall, but it doesn't tells us what feature of the stimulus are better encoded, nor which specific interactions between signal and noise allow for that benefit. 
We wondered what features of the signal were enhanced by strong positive noise correlations in our population encoding model.

Thanks to the translation-invariant structure of the model, the mutual information and noise synergies may be decomposed spectrally as a sum over spatial frequencies $k$ (expressed in units of inverse distance between nearest neighbors):
\begin{equation}\label{spectralnoisesynergy}
  \Delta I=\frac{n}{2}\int_{-1/2}^{1/2} {\rm d}k\, \log\left(\frac{1+S(k)/N(k)}{1+S(k)/V_{\rm n}}\right),
\end{equation}
where $S(k)$ is the power spectrum of the stimulus, and $N(k)$ that of the noise (see Methods), and $n\to\infty$ is the total number of neurons. 
In this decomposition, low frequencies correspond to long-range collective modes, while high frequencies correspond to fine-grain features.

Natural stimuli involve spatially extended features impacting many neurons. This causes neural responses to exhibit strong long-range signal correlations between neurons, corresponding in our model to large $L_{\rm s}$ (Fig.~\ref{fig:fig2}B). Most information is then carried by low frequency modes of the response (Fig.~\ref{fig:fig3}A).

Noise correlations concentrate noise power at low frequencies {\GM and decrease noise power at high frequencies} for a fixed noise level $V_{\rm n}$ (inset of Fig.~\ref{fig:fig3}B).
As a result, noise correlations enhance information in the high frequency modes of the signal ($k \geq k^*$), at the expense of the low frequencies features (Fig.~\ref{fig:fig3}B), which are already well represented.
Fig.~\ref{fig:fig3}C shows the spectral decomposition of the noise synergy as a function of the noise correlation range $L_{\rm n}$.
The critical frequency $k^*=(1/2\pi)\arccos(e^{-1/L_{\rm n}})$ above which noise correlations are beneficial only depends on $L_{\rm n}$ (Fig.~\ref{fig:fig3}C). However, the relative information gains in each frequency domain depends on the strengths of the signal and noise correlations.

In summary, noise correlations enhance fine details of the stimulus to the detriment of its broad features, which are already sufficiently well encoded. 
This redistribution of the noise across the spectrum drives the gain in information. {This effect is generic to any choice of the correlation lengths, and we expect it to hold for other forms of the power spectra and receptive field geometries.}

\subsection*{Noise correlations in the retina favor the encoding of fine stimulus details}

 % fig4
\begin{figure*}
\begin{center}
\includegraphics[width=1.\textwidth]{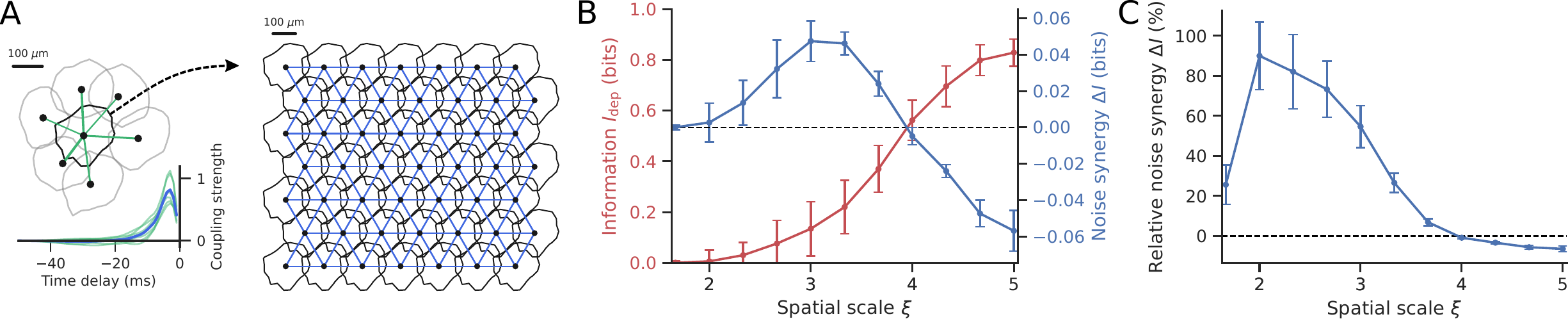}
\end{center}
\caption{
{\bf Noise correlations benefit small scale features at the detriment of large scale ones}.
{\bf A.} We built a large population of 49 RGCs based on 7 neurons recorded from the mouse retina. A deep-GLM \cite{Mahuas20} was fit to the experimental population and its central neuron model was tilled on a triangular lattice to create a large RGC network.
Couplings between the central experimental cell and its neighbors were symmetrized (green links in the population plot; green lines in the inset plot) and averaged to obtain the coupling filters between nearest neighbors in the synthetic population (blue links in the synthetic mosaic; blue line in the inset plot).
{\bf B.} {\TM Information $I_{\rm dep}$} and noise synergy $\Delta I$ per pixel for stimulus features of increasing scales {\TM ($\xi=2$ check sizes in units of inter-neuron distance)}. These quantities were computed via a decoding approach applied to a binary flashed checkerboard stimuli with various check sizes. Error bars are the standard error obtained by repeating the analysis on bootstrapped data.
In the absence of noise correlations, little information is transmitted about small stimulus features. By contrast, large scale features are well encoded and information per pixel saturates towards 1 bit as check size grows.
The noise synergy is positive for small and intermediate check sizes while negative for larger checks, in line with the theoretical results highlighted in Fig. \ref{fig:fig3}.
{\bf C.} Noise correlations nearly double the amount of information encoded about stimulus features of small and intermediate sizes, while only decreasing information for the largest checks by less than 10\%.
}
\label{fig:fig4}
\end{figure*}

To test our predictions, we studied experimentally the impact of noise correlations on the encoding of features at different spatial scales in the retina. We recorded {\em ex vivo} the spiking activity of 7 OFF-$\alpha$ retinal ganglion cells from a mouse retina using the same experimental technique as described before.
We presented the retina with a multi-scale checkerboard stimulus composed of frames made of random black and white checkers, flashed at 4\,Hz. Each frame was made of a checkerboard with a given spatial resolution (checks of sizes $12$, $24$, $36$, $72$ and $108$ $\mu$m).
From the recorded activity, we infered an {\TM artificial convolutional neural network (CNN) with interactions} \cite{Mahuas20} and {\GM used the inferred model to build a large synthetic population of $49$ cells organized on a triangular lattice (Fig.~\ref{fig:fig4}A). {\TM Previous studies have demonstrated the ability of CNN to capture the stimulus-response of RGCs acurately \cite{mcintosh2016deep,maheswaranathan2023interpreting,ding2023information,azeglio2025higher,goldin2022context}.}
  {\TM We checked that the resulting correlation structure was consistent with previous reports, following an exponential decay as a function of distance with a characteristic length of $110\mu$m (Supplemental Fig. S4), comparable with observations of 170-290 $\mu$m in the rat \cite{Ruda20}.}
  We then generated a large dataset of repeated responses to regular black and white checker flashes. Each checker was composed of checks of a given size (sizes ranging from $140$ to $420$ $\mu$m, with $28$ $\mu$m increments) and for each check size, $50$ spatially offset versions of the checker were showed. We trained a linear decoder of each pixel value (black or white) on this synthetic dataset, and a second decoder on the synthetic data in which the activity of each cell was shuffled across repetitions to destroy noise correlations (see Methods).

The two decoders were then applied to the testing datasets, synthetically generated in the same way as the training sets, to decode each pixel from the response.
For a fair comparison, the second decoder was applied to data in which noise correlations were removed by shuffling, as in the training.
The mutual information carried by the decoders was then estimated separately for each checker size.
To limit border effects, the mutual information was estimated for each pixel within a small hexagon centered on the central cell of the synthetic population, of size (distance between opposite sides of the hexagon) equal to the distance between cells.

We found that the gain in mutual information afforded by noise correlations is large and positive for small and intermediate check sizes, while moderately negative for large checks (Fig.~\ref{fig:fig4}B and C).
These results suggest that noise correlation benefit the encoding of small-scale features of the stimulus, at the expense of the large-scale ones, which are easier to encode. 
Noise correlations can therefore trade the encoding power of large-scale features to improve sensitivity to the small-scale ones.}

\section{Discussion}

 % discussion
Many experimental works have shown that neurons with the strongest positive noise correlations are similarly tuned to the stimulus \cite{Mastronarde83,Mastronarde89,Zohary94,Lee98,Bair01,Kohn05,Gutnisky08,Averbeck06,Hofer11}.
Here the sign rule \cite{Averbeck2006review,Hu14,Azeredo21} would predict a detrimental effect of shared variability, {at odds with the efficient coding hypothesis \cite{Barlow61}, which is supported by a large body of work showing that noise correlations are indeed beneficial
\cite{Abbott99,Pillow08,Ecker11,Azeredo14,Franke16,Ruda20}.}
Our work resolves this inconsistency by showing that beyond a critical value $\rho_{\rm n}^*$, noise correlations become beneficial to information encoding regardless of their sign.
We experimentally demonstrated this effect in recordings of retinal neurons subject to stimuli with different statistics, and showed that it generalizes to large populations of sensory neurons. {\TM The effect depends on the spatial scale}: large scale (low dimensional) modes give rise to strong signal correlations, making positive noise correlations detrimental, while small scale (high dimensional) modes benefit from positive noise correlations since their signal correlations are small. {\TM These small-scale features correspond to high frequencies, and thus to local differential features, consistent with the analysis on pairs of neurons showing that accuracy in the difference of neural activities is enhanced by noise correlations (Fig.~\ref{fig:fig1}I).}

{\TM Since much theoretical work has been done on the impact of noise correlations, it is useful to emphasize the different ways in which our approach differs from previous studies. First, we consider non-linear effects in the noise parameter (Eq.~\ref{eq:dMI_2cd}). Crucially, the second-order term is responsible for the violation of the sign rule. Second, we study mutual information rather than Fisher information. However, most of our results also hold for the Fisher information, see Supplemental Material, Appendix B. Third, we used constant, stimulus-independent noise correlations, in the sense of Pearson. However, we also showed that, counter-intuitively, strong noise correlations could be beneficial even in the case in ``information-limiting correlations'' \cite{Moreno-Bote14, Kanitscheider15} (Supplemental Material, Appendix A and Fig. S1), which is the most defavorable case of stimulus-dependent noise correlation. More generally, we considered the impact of such stimulus-dependent noise correlations within our framework (see Supplemental Material, Appendix A), and showed that these fluctuations can improve the noise synergy in two ways: by being large, and by being {\TM correlated with} the noise level $V_{\rm n}(\theta)$, also assumed to be stimlulus dependent. Our results thus extend and clarify previous theoretical work  \cite{Panzeri99a, Pola03} under a common information-theoretic framework.
Lastly, relative to \cite{Sompolinsky01} we take the large population limit by considering large networks with finite correlation length, in units of inter-neuron separation, rather than by increasing the density of neurons; that scaling makes it possible to uncover the benefit of noise correlations at short length scales.}

{\TM Our spectral analysis of scales and dimensionality helps understand apparently contradictory claims in the literature. On the one hand,} noise {\TM correlations} should be detrimental for coding, because it impedes denoising by pooling the signal of multiple neurons \cite{Zohary94,Sompolinsky01}.
{\TM On the other hand,} studies focusing on the stimulus response of large sensory populations have observed a positive gain \cite{Pillow08,Ecker11,Graf11,Franke16,Ruda20}.
Our study {\TM suggests the following interpretation:}
when the neural population encodes a low dimensional stimulus, as the angle of a drifting gratings, similarly tuned nearby neurons become strongly signal-correlated, and their noise correlations are detrimental \cite{Sompolinsky01}.
In the case of high dimensional stimuli, like naturalistic images or videos, signal correlations between them are positive but weak, so that noise correlations become larger than the threshold $\rho_n^*$, and therefore beneficial.

{Our work also shares some similarity with Ref.~\cite{Tkacik10}, where the authors predicted the optimal patterns of noise correlations maximizing information transmission by a population of neurons. They showed that at high SNR, optimal noise correlations follow the sign rule. This result does not rule out that high levels of noise correlations violating the sign rule could be beneficial---albeit not optimal---in agreement with our theory. However, a direct comparison with our results is difficult because in Ref.~\cite{Tkacik10} noise correlations were tuned through inter-neuron couplings that affect the mean response of each neuron to the stimulus, which is kept constant in our analysis. In fact, this effect leads to optimal noise correlations of the same sign as signal correlations in the low SNR regime \cite{Tkacik10}. It was also shown to improve positional coding in the hippocampus through the sharpening of stimulus tuning \cite{Nardin23}. This apparent violation of the sign rule is however indirect and distinct from the direct beneficial effect of strong noise correlations that we discuss in this work.}

Several studies have focused on the effect of noise correlations on the Fisher information \cite{Sompolinsky01,Ecker11,Franke16,Abbott99,Smith08}.
While our main results are based on the mutual information, {they equivalently apply to the Fisher information in the Gaussian case \cite{Brunel98} (see Supplemental Material, Appendix B)}.
To further test the robustness of our conclusions, we demonstrated that our results are model independent, and hold both for binary and Gaussian neurons. In addition, empirical results from the retinal recordings (Fig.~\ref{fig:fig1}J) were obtained without any approximation or model choice, and agree with the theory.

We validated our theoretical predictions experimentally on recordings of neurons from the retina.
{\TM Our theory could be further tested in large-scale retinal recordings in which the population size is large enough to observe network effects, by presenting stimuli in which signal correlations are weak or tunable. An interesting question is whether the magnitude of signal correlations in natural scenes is consistent with a beneficial effect of noise correlations, in line with arguments of efficient coding in the retina \cite{Atick92,Pitkow12}.}
Applying our approach to data in sensory cortical areas where similar noise correlation structures have been observed \cite{Hofer11,Smith08} could lead to new understanding of the role of noise correlations in sensory information processing.
Another key open question is what stimulus ensembles most benefit from noise correlations, and where naturalistic stimuli stand in that regard.
We have further shown that noise correlations benefit the encoding of high-frequency features of the stimulus, which correspond to fine-grained neural activity patterns. Extending these results to higher cortical areas would require understanding which features from the stimulus drive such activity patterns.

\section{Methods}

 % methods
\subsection*{Covariance and correlation measures}
The average responses of two neurons 1 and 2 are given as function of the stimulus $\theta$ by the tuning curves $\mu_1(\theta) = \langle r_1 \rangle_{\theta}$ and $\mu_2(\theta) = \langle r_2 \rangle_{\theta}$.
Signal correlations are defined as
$\rho_{\rm s} = \mathrm{Corr}_\theta(\mu_1,\mu_2)$,
and noise correlations as:
$\rho_{\rm n}(\theta) = \mathrm{Corr}(r_1,r_2|\theta)$.
The sum of these two coefficients does not have a simple interpretation in terms of total correlation or covariance, but we can also decompose the total correlation coefficient betweeen $r_1$ and $r_2$ as $\mathrm{Corr}(r_1,r_2) = r_{\rm s}+r_{\rm n}$, with
$r_{\rm s} = {\mathrm{Cov}_\theta(\mu_1,\mu_2)}/{\sqrt{\mathrm{Var}(r_1)\mathrm{Var}(r_2)}}$,
and
$r_{\rm n} = {\langle\mathrm{Cov}(r_1,r_2|\theta)\rangle_{\rm \theta}}/{\sqrt{\mathrm{Var}(r_1)\mathrm{Var}(r_2)}}$.

\subsection*{Pairwise analysis}
\paragraph*{Tuning curves.}
We consider a pair of neurons encoding an angle $\theta$. The responses of the two neurons, $r_1$ and $r_2$, are assumed to be binary (spike or no spike in a 10\,ms time window) and correlated. 
Their average responses $\mu_1(\theta)$ and $\mu_2(\theta)$ are given by Von Mises functions (Fig. \ref{fig:fig1}A):
\begin{equation} \label{eq:VonMises}
\mu_i(\theta) = a\frac{\exp{\left(\cos{\left( \theta - \theta_{\rm c}^i \right)/w}\right)} - \exp{\left( -1/w \right)} }{ \exp{\left(1/w\right)} - \exp{\left(-1/w\right)} } + b.
\end{equation}
Signal correlations between the two neurons can be tuned by varying the distance between the center of the two tuning curves $\theta_{\rm c}^1$ and $\theta_{\rm c}^2$. {\GM The tuning curve width $w$ was set arbitrarily to $5$, the amplitude $a$ to $0.4$ and the baseline $b$ to $0.1$.} The strength of noise correlations is set to a constant of $\theta$, $\rho_{\rm n}(\theta)=\rho_{\rm n}$.

\paragraph*{Small correlation expansion.}
When noise correlations $\rho_n$ are constant and small, the noise synergy may be expanded as \cite{Mahuas23}:
\begin{equation}\label{synergybinarypair}
\Delta I {\UF \approx} {-r_{\rm s} r_{\rm s} + \frac{1}{2}\left( {\rho_{\rm n}}^2 - {r_{\rm n}}^2 \right)}=\frac{\alpha}2 \rho_{\rm n} \left( \rho_{\rm n} - \rho_{\rm n}^* \right),
\end{equation}
where the second equality highlights the dependency on $\rho_n$. The critical $\rho_n^*$ may be written as
\begin{equation}\label{eq:rhostarmeth}
\rho_{\rm n}^* = \beta \, \rho_{\rm s},\quad\textrm{with}\quad \beta=\frac{2V_{\rm s} V_{\rm n}}{{V_{\rm tot}}^2-{V_{\rm n}}^2},
\end{equation}
and the prefactor $\alpha = 1 - V_{\rm n}^2/V_{\rm tot}^2$,
with the shorthands $V_{\rm tot}=\sqrt{\mathrm{Var}(r_1) \mathrm{Var}(r_2)}$, $V_{\rm n}=\<\sqrt{\mathrm{Var}(r_1|\theta) \mathrm{Var}(r_2|\theta)}\>_\theta$, and $V_{\rm s}=\sqrt{\mathrm{Var}(\mu_1(\theta)) \mathrm{Var}(\mu_2(\theta))}$ corresponding to measures of total, noise, and signal variances in the two cells.

By Cauchy-Schwartz we have:
\begin{equation}
{V_{\rm n}}^2 \leq \<\mathrm{Var}(r_1|\theta)\>_\theta \<\mathrm{Var}(r_2|\theta)\>_\theta,
\end{equation}
which entails
\begin{equation}\label{eq:betabinarbound}
\beta \lesssim \frac{1}{\cosh \frac{\Delta\textrm{ln}R}{2}+R/2} \leq 1,
\end{equation}
where $R=\sqrt{R_1 R_2}$ and $\Delta\textrm{ln}R=\ln(R_1/R_2)$, with $R_i=\mathrm{Var}(\mu_i)/\<\mathrm{Var}(r_i|\theta)\>_{\theta}$ the signal-to-noise ratio of the cells. $R$ measures the overall strength of signal-to-noise ratios, while {\TM $\Delta\textrm{ln}R$} measures their dissimilarity. The last inequality implies that noise correlations are always beneficial for $\rho_{\rm n}>\rho_{\rm s}$.

\paragraph*{Gaussian case.}
To test the theory's robustness to modeling choices, we also considered a continuous rather than binary neural response: $r_i=\mu_i(\theta)+\delta r_i$, where both $\mu_i$ and $\delta r_i$ are Gaussian variables defined by their covariance matrices $\Sigma_{{\rm s},ij}=\mathrm{Cov}_\theta(\mu_i,\mu_j)$, and $\Sigma_{{\rm n},ij} = \langle\mathrm{Cov}(r_i,r_j|\theta)\rangle_{\rm \theta}$.
The noise synergy can be calculated through classic formulas for the entropy for Gaussian variables, yielding:
\begin{equation}\label{eq:GaussiandMI}
\Delta I = \frac{1}{2}\log{\left(\frac{|\Sigma_{\rm s} + \Sigma_{\rm n}||V_{\rm n}|}{|\Sigma_{\rm s} + V_{\rm n}||\Sigma_{\rm n}|}\right)},
\end{equation}
where $|X|$ denotes the determinant of matrix $X$, and where $V_{\rm n}$ is the diagonal matrix containing the noise variances  of the cells $V_{{\rm n},ii}=\Sigma_{{\rm n},ii}$. Note that this formula is general for an arbitrary number of correlated neurons. In the pairwise case considered here matrices are of size $2\times2$.
The condition for beneficial noise correlations $\Delta I\geq 0$ is satisfied for $\rho_{\rm n}\geq \rho_{\rm n}^*$, with
\begin{equation}
\rho_{\rm n}^* = \beta \rho_{\rm s} \quad\textrm{with } \beta=\frac{1}{\cosh \frac{\Delta\textrm{ln}R}{2} + (1-\rho_s^2) R/2}\leq 1,
\end{equation}
which has a similar form as Eq.~\ref{eq:betabinarbound}.

\paragraph*{Noise synergy at constant noise entropy.}
Increasing noise correlations at constant $V_{\rm n}$ decreases the effective variability of the response, as measured by the noise entropy, $H(\{r_1,r_2\}|\theta)=\ln(2\pi e|\Sigma_{\rm n}|^{1/2})$, with $|\Sigma_{\rm n}|=V_{\rm n}^2(1-\rho_{\rm n}^2)$ in the case two neurons with the same noise level. To correct for this effect we also computed $\Delta I$ at constant noise entropy, by rescaling the noise variances in the correlated and uncorrelated cases,  $V_{\rm n,c}$ and $V_{\rm n,u}$, so that their resulting noise entropies are equal $|\Sigma_{\rm n}|=V_{\rm n,c}^2(1-\rho_{\rm n}^2)=V_{\rm n,u}^2$.

The critical noise correlation at which $\Delta I\geq 0$ is then given by:
\begin{equation}
\rho_{\rm n}^* = 2 \frac{\rho_{\rm s}}{1+\rho_{\rm s}^2}\leq 1,
\end{equation}
where the last inequality implies that strong enough noise correlations are always beneficial.

\paragraph*{Retinal data.}
Retinal data were recorded ex-vivo from a rat retina using a microelectrode array \cite{Deny17} and sorted using SpyKING CIRCUS \cite{Yger18} to isolate single neuron spike trains. From the ensemble of single cells we could isolate a population of 32 OFF-$\alpha$ ganglion cells. 
Three stimuli movie with different spatio-temporal statistics were presented to the retina: a checkboard movie consisting of black and white checks changing color randomly at 40\,Hz and repeated $79$ times; a drifting grating movie consisting of black and white stripes of width $333\,\mu$m moving in a fixed direction relatively to the retina, at speed $1$\,mm/s, and repeated $120$ times; and finally a movie composed of $10$ black disks jittering according to a Brownian motion on a white background, repeated $54$ times.

\subsection*{{\GM Gaussian population and} spectral analysis}
We consider a population of $n$ neurons organized along a 1D lattice with constant interneuron spacing. Their mean response and noise are assumed to be Gaussian, with their noise and signal covariances given by an exponentially decaying function of their pairwise distances:
\begin{align}
&\Sigma_{{\rm s},ij} = V_{\rm s} e^{-|i-j|/L_{\rm s}}, \\
&\Sigma_{{\rm n},ij} = V_{\rm n}(\delta_{ij}+\rho_{\rm n}^{\rm 0} e^{-|i-j|/L_{\rm n}}).
\end{align}
$V_{\rm s}$ and $V_{\rm n}$ are the signal and noise variance of the single cells. The parameter $\rho_{\rm n}^{\rm 0}$ sets the strength of noise correlations such that nearest neighbors have noise correlation $\rho_{\rm n} \equiv \rho_{\rm n}^0 \exp{\left( -1/L_{\rm n} \right)}$. When $n$ is large and boundary effects can be ignored, the system is invariant by translation and we can diagonalize $\Sigma_{\rm s}$ and $\Sigma_{\rm n}$ in the Fourier basis $\nu_{k,l} = \frac{1}{\sqrt{n}}\exp{\left( - i 2 \pi k l / n \right)}$. Denoting the spectra of $\Sigma_{\rm s}$ and $\Sigma_{\rm n}$ by $S(l/n)$ and $N(l/n)$,
the expression of the noise synergy, Eq. \ref{eq:GaussiandMI}, can then be written as a sum over modes:
\begin{equation}\label{eq:deltaI_decomposition}
\Delta I = \frac{1}{2}\sum_{l=-\frac{(n-1)}{2}}^{\frac{(n-1)}{2}}\log{\left( \frac{1+S(l/n)/N(l/n)}{1+S(l/n)/V_{\rm n}} \right)},
\end{equation}
which simplifies in the $n\to\infty$ limit to:
\begin{equation}\label{eq:deltaI_decomposition_cont}
\frac{\Delta I}{n} = \frac{1}{2}\int_{-1/2}^{1/2}\log{\left( \frac{1+S(k)/N(k)}{1+S(k)/V_{\rm n}} \right) {\rm d}k},
\end{equation}
with
\begin{align}
&S(k) = V_{\rm s} \frac{1-{\rho_{\rm s}}^2}{1-2 \rho_{\rm s} \cos{\left(2 \pi k  \right)} + {\rho_{\rm s}}^2}, \label{eq:signal_eigvals} \\
&N(k) = V_{\rm n} \left( 1 - \rho_{\rm n}^0 +\rho_{\rm n}^0 \frac{1-{\lambda_{\rm n}}^2}{1-2 \lambda_{\rm n} \cos{\left(2 \pi k \right)} + {\lambda_{\rm n}}^2} \right) \label{eq:noise_eigvals} ,
\end{align}
where $\rho_{\rm s} = \exp{\left( -1/L_{\rm s} \right)}$ is the nearest-neighbors signal correlation, and $\lambda_{\rm n} = \exp{\left( -1/L_{\rm n} \right)}$. $k$ is a wave vector interpretable as a spatial frequency in units of the system's size, up to a $2\pi$ factor. Examining Eq.~\ref{eq:deltaI_decomposition_cont}, we see that noise correlations are beneficial for frequencies for which $N(k) \leq V_{\rm n}$, which happens for $k \geq k^*$ where $k^* = (1/2\pi)\arccos(e^{-1/L_{\rm n}})$.

In the low noise regime, $R = V_{\rm s}/V_{\rm n} \gg 1$, the noise synergy reduces to:
\begin{equation}
\frac{\Delta I}{n}  \approx  -\frac{1}{2}\int_{-1/2}^{1/2}\log{\left( N(k)/V_{\rm n} \right)}dk\geq 0,
\end{equation}
where the inequality stems from Jensen's inequality, because $-\log$ is a convex function, and $-\log(\int_{-1/2}^{1/2}dk N(k)/V_{\rm n})= 0$. Therefore in that regime noise correlations are always beneficial.

In the high noise limit, $R \ll 1$, the noise synergy becomes:
\begin{equation} \label{eq:deltaI_highnoise}
\frac{\Delta I}{n} \approx \frac{1}{2} \int_{-1/2}^{1/2} \left[\frac{S(k)}{N(k)} - \frac{S(k)}{V_{\rm n}}\right] dk.
\end{equation}
Computing this integral gives the critical noise correlation:
\begin{align}\label{eq:rhostar_gaussian}
\rho_{\rm n}^* = \rho_{\rm s}\frac{(1 - \lambda_{\rm n}^2)}{1 - 2 \lambda_{\rm n}\rho_{\rm s}+{\rho_{\rm s}}^2}\leq \rho_{\rm n}^{\rm max},
\end{align}
where $\rho_{\rm n}^{\rm max}=(1+\lambda_{\rm n})/2$ is the maximum possible value of $\rho_{\rm n}$ (ensuring that the noise spectrum $N(k)$ is non-negative for all $k$). The last inequality in Eq.~\ref{eq:rhostar_gaussian} implies that there always exists a regime in which strong noise correlations are beneficial.

\paragraph*{Non-identical neurons.}
To study the effect of nonhomogeneities among neurons, we considered the case where the signal variance of each cell is different, and drawn at random as $\sqrt{V_{\rm s}^i} = \mu + \eta_i$, where $\eta_i$ is normally distributed with zero mean and variance $\nu^2$.
The noise synergy can be rewritten in the high noise regime ($R\ll 1$) as:
\begin{equation}
\Delta I \approx \frac{1}{2}\mathrm{Tr}\left(\Sigma_{\rm s}\Sigma_{\rm n}^{-1} - \Sigma_{\rm s}V_{\rm n}^{-1}\right).
\end{equation}
Averaging this expression over $\eta_i$ yields:
\begin{equation} \label{eq:dIinhomogeneous}
\Delta I \approx {\Delta I}_{\rm u} + \frac{1}{2}\frac{\nu^2}{(\mu^2+\nu^2)}\mathrm{Tr}{\left( \Sigma_{\rm n}^{-1}\left( \bar{R}\mathbb{I} - \Sigma_{\rm s} \right) \right)},
\end{equation}
where $\Delta I_{\rm u}$ is the noise synergy in a uniform population (with $V_{\rm s}=\mu^2+\nu^2$), and
where {\GM the second term is always positive, with} $\bar{R}=\langle V_{\rm s}^i \rangle/V_{\rm n} = (\mu^2 + \nu^2)/V_{\rm n}$.

Taking the continuous limit ($n\to\infty$) in Eq. \ref{eq:dIinhomogeneous}, similarly to the integral limit of Eq.~\ref{eq:deltaI_decomposition_cont}, allows us to write the critical noise correlation $\rho_{\rm n}^*$ as:
\begin{equation}
\rho_{\rm n}^* = \frac{\rho_{\rm n}^{*,{\rm u}}}{1+\gamma} + \frac{1-\rho_{\rm n}^{*,{\rm u}}}{2}\left(1-\sqrt{1+\frac{4 \gamma \rho_{\rm s}^2 (1-\lambda_{\rm n}^2)}{(1+\gamma)^2(1-\lambda_{\rm n}\rho_{\rm s})^2}} \right),
\end{equation}
where $\gamma = \nu^2/\mu^2$ quantifies the relative magnitude of nonhomogeneities, and $\rho_{\rm n}^{*,{\rm u}}$ is the critical noise correlation value in a uniform population (Eq.~\ref{eq:rhostar_gaussian}). This modified critical noise correlation value is always smaller than in the uniform case, and scales linearly  at leading order with the inhomogeneity parameter $\gamma$:
\begin{equation}
\rho_{\rm n}^* = \rho_{\rm n}^{*,{\rm u}} \left( 1-\gamma \frac{1-\rho_{\rm s}^2}{1-\rho_{\rm s}\lambda_{\rm n}} \right) + o(\gamma).
\end{equation}

\subsection*{Decoding analysis}
\paragraph*{Experimental and synthetic data.}
{\GM
We presented a mouse retina with a stimulus consisting of a black and white random checkerboard flashed at 4Hz, each frame with a given spatial resolution (checks of sizes 12, 24, 36, 72 and 108\,$\mu$m). Retinal ganglion cell activity was recorded ex-vivo using a micro-electrode array and single neuron activity isolated via spike sorting using SpyKING CIRCUS \cite{Yger18}. We isolated a population of $N_{\rm cells}=7$ OFF-$\alpha$ retinal ganglion cells which presents strong noise correlations in their response \cite{Volgyi13}.
The original recording contained a $15\,s$ checkerboard movie repeated 90 times as well as 90 different $22.5\,s$ long unrepeated movies.

We inferred a deep Generalized Linear Model (GLM) of the central cell
among 7 from the experimental population (Fig.~\ref{fig:fig4}A),
consisting of a stimulus-processing filter, and filters for the spiking history of
the cell as well as its 6 neighbors (couplings).
The stimulus-processing part of the model consisted of a deep neural network composed of two spatio-temporal convolutional layers followed by a readout layer. The whole model was fit to the data using the 2-step inference approach \cite{Mahuas20}. 

A synthetic population of $49$ OFF-$\alpha$ ganglion cells was then constructed by arranging them on a triangular lattice of $7$ by $7$ points. Each cell responds according to the inferred GLM with translated receptive fields. Nearest neighbors were coupled with the average of the GLM couplings inferred between the central cell from the experiment and its neighbors.

To stimulate this synthetic population, we generated a synthetic stimulus ensemble from $550$ regular black and white checker frames, each with a given check size ranging from $140$ to $420$ $\mu$m (with increments of $28$ $\mu$m). Every checker of a given size was presented for $5$ different regularly spaced offsets ranging from $0$ to $224$ $\mu$m both in the horizontal and vertical directions, resulting in $25$ different frames per size. To further ensure that the color of each pixel in the stimulus ensemble is black or white with equal probability, each checker frame also had its color-reversed version in the set, resulting in $50$ different frames for a given check size. A single snippet from the synthetic stimulus ensemble consisted of a $250$ms white frame followed by one of the $550$ aforementioned checker frames.

We built a training, a vadliation, and testing set for the dependent and independent decoders by simulating the synthetic population for sets of 3750, 1250, and 5000 repetitions (respectively) of each synthetic stimulus snippets.
}

\paragraph*{Decoders.}
{\GM
The binary decoders are logistic regressors taking in the integrated response of the population over the $N_{\tau} = 5$ past bins of $50$\,ms to predict the ongoing stimulus frame. The predicted stimulus at time $t$ and repeat $k$ is given by:
\begin{equation}
\hat{X}(x,y,t,k) = f\left( A_{x,y} r(t,k) + \beta_{x,y} \right),
\end{equation}
where $x,y$ are the pixel indices along the two dimensions of the stimulus, $f(x)=(1+e^{-x})^{-1}$ is the sigmoidal function, $A_{x,y}$ is a matrix of size $(N_{\tau},N_{\rm cells})$, $r(t,k)$ is a matrix of size $(N_{\rm cells},N_{\tau})$ containing the spike history of the population at time $t$ and repeat $k$, and $\beta_{x,y}$ a pixel-wise bias.
Each decoder was trained by minimizing the average binary cross entropy (BCE) between predicted stimulus $\hat{X}(x,y,t,k)$ and the true stimulus $X(x,y,t)$, $\langle \mathrm{BCE}(x,y,t,k)\rangle_{x,y,t,k}$, where
\begin{equation}\label{BCE_train}
\begin{split}
  \mathrm{BCE}(x,y,t,k)= - X(x,y,t)\ln{(\hat{X}(x,y,t,k))}\\
  - (1-X(x,y,t))\ln{(1-\hat{X}(x,y,t,k))}.
\end{split}
\end{equation}
Training was done by stochastic gradient descent on the synthetic datasets using the training (3750 repetitions) and validation (1250 repetitions) sets. Optimization was done using stochastic gradient descent with momentum, with early stopping when the validation loss did not improve over 6 consecutive epochs. During that procedure the learning rate was divided by 4 whenever the validation loss did not improve for 3 consecutive epochs.

We probed the decoders' abilities to decode features of different spatial scales by decoding the simulated responses of the synthetic population to the checker stimuli with varying check size from the testing set.
Performances of the decoders were assessed by computing the mutual information between each pixel's color $X$ and it's decoded value $\hat X$, separately for the different sizes of checks. The noise synergy was then computed as the difference between the mutual information averaged over pixels for the dependent and independent decoders.

Error bars were computed as follows. We infered $10$ deep GLMs on bootstraps of the original training set, obtained by re-sampling with replacement the simulus-response pairs used for training. These $10$ models were used to generate $10$ surrogate training sets, from which $10$ separate decoders were infered with noise correlations, and another $10$ without noise correlations.
Then synthetic test sets for the checker decoding task were generated from each of the $10$ models, and the performance of each decoder computed separately with and without noise correlations, yielding $10$ values of the mutual information, and $10$ values of the noise synergy (both averaged over pixels). The error bars are the standard deviations of the resulting information, noise synergy and synergy-to-information ratios (i.e. relative noise synergy) over the $10$ bootstraps.
}

\section*{Data availibility}
Part of the data utilized in this work have been used or published in previous studies \cite{Deny17, Botella18}. The remaining data and codes will be shared upon publication of this study. The code used to generate the synthetic data is available at \url{https://github.com/gmahuas/noisecorr}.

\section*{Acknowledgements}
 % thanks
We thank St\'ephane Deny for help with the experimental data used in the paper and Matthew Chalk and Simone Azeglio for useful discussions. This work was supported by ANR grants n. ANR-22-CE37-0023 ``LOCONNECT'' and n. ANR-21-CE37-0024 NatNetNoise, and by IHU FOReSIGHT (ANR-18-IAHU-01) and by Sorbonne Center for Artificial Intelligence- Sorbonne University- IDEX SUPER 11-IDEX-0004. This work was also supported by the Bettencourt Schueller Foundation.

\bibliographystyle{pnas}

\onecolumngrid

\newpage

\appendix

 % si
\renewcommand{\thefigure}{S\arabic{figure}}
\setcounter{figure}{0}

\section{Stimulus-dependent correlations}  \label{SI:S1}
Theoretical developments in the current work focused on the effect of \textit{constant} noise correlation {(in the sense of Pearson)} on the mutual information between stimulus and response. In reality, noise correlations, i.e. correlations of the response conditioned on stimulus, may also depend on the stimulus value. This dependency may arise either via a dependence to the firing rates of the cells \cite{Delarocha07} or to the stimulus itself \cite{Franke16, Trenholm14}. Although a complete discussion of the effect of non-constant noise correlations on mutual information is out of this paper's scope, we succintly extend our framework to give intuition about this phenomenon.

Starting again from the framework described in section \textit{Pairwise analysis} of the methods and relaxing the assumption of constant noise correlations we can show (Eq. A7 in ref. \cite{Mahuas23}) that the noise synergy expanded and truncated at second order in the correlations for a pair of cells encoding for stimulus $\theta$ is given, with the definitions from the main text Methods section, by:
\begin{equation}
\Delta I = -r_{\rm s}r_{\rm n} + \frac{1}{2}\left( \langle {\rho_{\rm n}}^2(\theta) \rangle_{\mathrm{\theta}} - {r_{\rm n}}^2 \right)
\end{equation}
We can decompose this information gain as:
\begin{equation}\label{eq:dMI_2cd_fluctuations}
\Delta I = \Delta I_{\rm c}+\Delta I_{\rm f,1} + \Delta I_{\rm f,2},
\end{equation}
where ${\Delta I}_c$ is given by Eq.~\ref{synergybinarypair} with constant noise correlations $\bar{\rho}_{\rm n} = \langle \rho_{\rm n}(\theta) \rangle_{\rm \theta}$, and the two other terms contain the effects of the variations of $\rho_{\rm n}(\theta)$. The first term $\Delta I_{\rm f,1}=(1/2)\mathrm{Var}_{\theta}(\rho_{\rm n}(\theta))\geq 0$ {directly} accounts for the effect of fluctuations of $\rho_{\rm n}(\theta)$, while $\Delta I_{\rm f,2}$ is linked to its correlation with the noise variance:
\begin{equation}
  \Delta I_{\rm f,2} = -
  \left\langle \left( \rho_{\rm n}(\theta) - \bar\rho_{\rm n} \right) \frac{V_{\rm n}(\theta)}{V_{\rm tot}} \right\rangle_{\theta} \times \\
  \left( \frac{1}{2} \left\langle \left( \rho_{\rm n}(\theta) + \bar\rho_{\rm n} \right) \frac{V_{\rm n}(\theta)}{V_{\rm tot}} \right\rangle_{\theta} + r_{\rm s} \right).
\end{equation}
This contribution is zero when noise correlations $\rho_{\rm n}(\theta)$ are not correlated to the noise variance of the pair $V_{\rm n}(\theta)$, but it can in general be positive or negative depending how these two quantities co-vary.
We saw in the analysis showed Fig.~\ref{fig:fig1} that the effect of strong positive noise correlations between OFF-$\alpha$ from the rat retina can be qualitatively explained by the second order approximation with constant noise correlations. This suggests that, at least for this type of neurons, the contribution of noise correlation fluctuations to the mutual information doesn't constitute the bulk of the effect.

Previous work investigated the effect of non-constant noise correlations on sensory coding. When focusing on the mutual information \cite{Pola03, Panzeri99a}, studies essentially relied on decompositions of the mutual information derived from correlation measures that stray away from classical Pearson correlations and their intuitive meaning (see App. D. in ref. \cite{Mahuas23} for a discussion on that point). As a consequence, interpretation of their results in terms of pairwise noise correlations is not straightforward.

Another line of work derived the notion of ``information-limiting'' correlations \cite{Moreno-Bote14,Kanitscheider15}, which are correlations that always align noise with the signal's direction, in any point of the manifold. Such correlations, however, are only truly information-limiting in the case of the Fisher information (they where in fact derived from the linear approximation of the Fisher information \cite{Moreno-Bote14}).
The Fisher information which only accounts for how an ideal decoder can distinguish two infinitely close stimuli $\theta$ and $\theta + {\rm d}\theta$ {with vanishing ${\rm d}\theta$. It thus relies on a linear analysis (first order in a small error parameter) and ignores higher-order effects.}
In contrast, the mutual information is a \textit{global} measure of information, that accounts for how the conditional response distributions are distinguishable from the marginal response distribution, and includes all orders of error parameter. While information-limiting correlation are always detrimental to the Fisher, they can benefit the mutual information by decreasing the overlap between conditionals. {Fig.~S1 A illustrates an example of such effect, in which the high curvature of the response curve causes information-limiting correlations to actually increase the discriminability of nearby stimuli. Since it is a function of the curvature, it is non-linear effect that cannot be captured by the Fisher information, but it does impact the mutual information.}

Fig.~S1 B quantifies the effect on the mutual information in a concrete case with Von Mises tuning curves (see caption for details). It shows that strong information-limiting correlations are always beneficial to the mutual information, while being always detrimental to the Fisher information (Fig.~S1 C). One must note however, that in the particular case of perfectly stimulus correlated neurons ($\rho_{\rm s} = 1$), information limiting correlations will always be detrimental, as they will increase the overlap between conditional response distributions and the marginal response distribution.

 % figS1
\begin{figure*}
\begin{center}
\includegraphics[width=1.\textwidth]{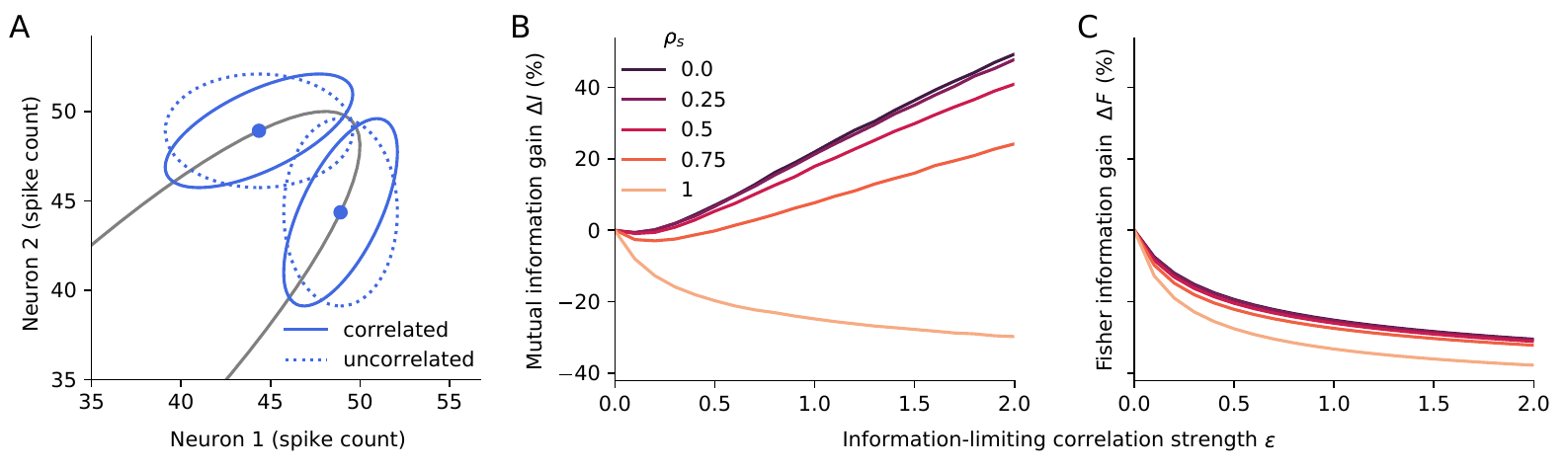}
\end{center}
\caption{
{\bf Impact of ``information-limiting correlations'' on stimulus information.}
An angular stimulus $\theta$ is encoded by a pair of neurons characterized by Von Mises tuning curves (with parameters $a=40$, $b=10$, and $w=5$).
Their response is Gaussian of means $\mu_1(\theta)$ and $\mu_2(\theta)$.
Information-limiting correlations are defined by a covariance of the form: $\Sigma_{\rm n}(\theta) =V_{0}\mathbb{I}+ \epsilon \mu'(\theta)\mu'(\theta) ^\top$, were $\epsilon$ controls their strength, and where we set $V_0=V_{\rm s}/2$. Note that $\Sigma_{\rm n}$ now depends on $\theta$.
{\bf A.} Noise covariance ellipses for the correlated (plain line) and uncorrelated (doted line) cases, for two example stimuli (blue points) on the manifold (grey), for an example of the system described above (with $\rho_{\rm s}\approx 0.9$). Here, information limiting correlations will decrease the overlap between the two conditional response distributions. The mutual information accounts not only for the detrimental local effects of information-limiting correlations, but also for such potentially beneficial {non-linear and global} effects.
{\bf B.} Relative mutual information gain (noise synergy) $\Delta I = I_{\rm dep}/I_{\rm indep} - 1$ as a function of $\epsilon$, where $I_{\rm dep}$ and $I_{\rm indep}$ quantify the mutual information with and without (off diagonal terms of $\Sigma_{\rm n}$ set to 0) noise correlations, for different levels of signal correlation $\rho_{\rm s}$. Mutual information was computed via Monte-Carlo integration.
Information-limiting noise correlations become beneficial to the mutual information if they are strong enough, except when cells are perfectly signal-correlated.
{\bf C.} By contrast, information-limiting correlations are always detrimental to the Fisher information $F(\theta)=\mu'(\theta) ^\top\Sigma_{\rm n}^{-1}(\theta)\mu'(\theta)$. The relative Fisher information gain $\Delta F = \langle (F_{\rm dep}(\theta)/F_{\rm indep}(\theta)- 1)\rangle_{\theta}$, where $F_{\rm dep}(\theta)$ and $F_{\rm indep}(\theta)$ denote the Fisher information with and without noise correlations, is always negative and decreases with $\epsilon$ and $\rho_s$.
}
\label{fig:figS1}
\end{figure*}

\section{Strong noise correlations and Fisher information}  \label{SI:S2}

We consider a pair of neurons encoding an arbitrary variable $\theta$. The responses $r$ of these neurons is assumed to be Gaussian {with } mean $\mu(\theta)$ and {constant} covariance matrix $\Sigma_{\rm n}$, where $\mu \left( \theta \right)$ are the tuning curves.
In this context the Fisher information is defined as \cite{Moreno-Bote14}:
\begin{equation}
 	F_{\rm dep}\left( \theta \right) = \mu'\left( \theta \right)^\top \Sigma_{\rm n}^{-1} \mu'\left( \theta \right).
\end{equation}
Expanding this expression for a pair of neurons with equal noise variance, $\Sigma_{\rm n}=V_{\rm n}\left(\begin{array}{cc} 1 & \rho_{\rm n} \\ \rho_{\rm n}&1 \end{array}\right)$, yields:
\begin{equation}
 	F_{\rm dep}\left( \theta \right) = \frac{{\mu'_{1}\left( \theta \right)}^2 + {\mu'_{2}\left( \theta \right)}^2}{V_{\rm n}\left( 1-\rho_{\rm n}^2 \right)} \left(1 - \rho_{\rm n} \frac{2\mu'_{1}\left( \theta \right)\mu'_{2}\left( \theta \right)}{{\mu'_{1}\left( \theta \right)}^2 + {\mu'_{2}\left( \theta \right)}^2}\right).
      \end{equation}
In the absence of noise correlations ($\rho_{\rm n}=0$), the Fisher information simplifies to:
\begin{equation}
 	F_{\rm indep}\left( \theta \right) = \frac{{\mu'_{1}\left( \theta \right)}^2 + {\mu'_{2}\left( \theta \right)}^2}{V_{\rm n}}.
\end{equation}
To quantify the overall Fisher improvement we introduce the quantity $\Delta F = \langle (F_{\rm dep}(\theta)/F_{\rm indep}(\theta)- 1)\rangle_{\theta}$. Defining $\xi(\theta) = \frac{2\mu'_{1}\left( \theta \right)\mu'_{2}\left( \theta \right)}{{\mu'_{1}\left( \theta \right)}^2 + {\mu'_{2}\left( \theta \right)}^2}$, the Fisher improvement becomes:
\begin{equation}
 	\Delta F = \frac{\rho_{\rm n}\left(\rho_{\rm n}- \langle\xi(\theta)\rangle_{\theta}\right)}{1 - {\rho_{\rm n}}^2}.
\end{equation}
Therefore, strong positive noise correlations will benefit the Fisher information whenever they exceed the critical value $\rho_{\rm n}^{*,\rm F} = \langle\xi(\theta)\rangle_{\theta}$.

 % figS2
\begin{figure*}
\begin{center}
\includegraphics[width=0.4\textwidth]{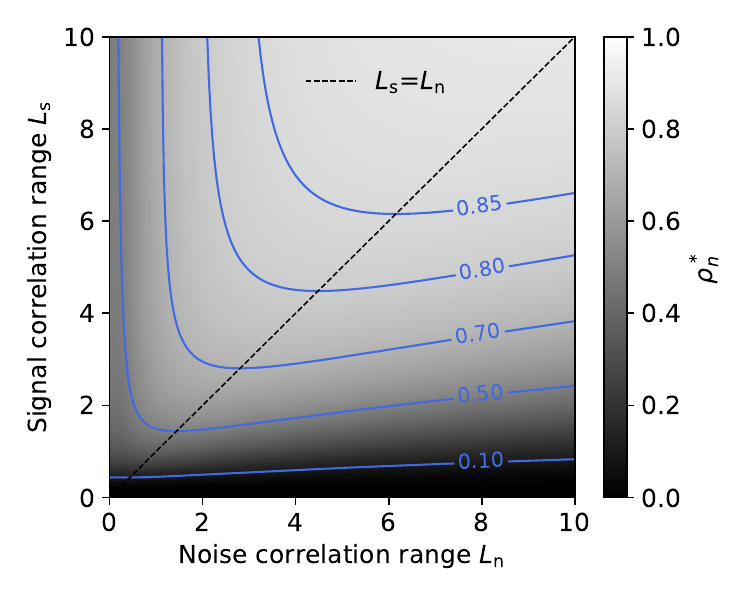}
\end{center}
\caption{
{\bf Behavior of $\rho_{\rm n}^*$}. $\rho_{\rm n}^*$ changes non-monotically with the signal $L_{\rm s}$ and noise $L_{\rm n}$ correlation ranges, and is concave with respect to these parameters. The maximum value of $\rho_{\rm n}^*$ {\GM at a given $L_{\rm s}$} is achieved when $L_{\rm n}=L_{\rm s}$.
}
\label{fig:figS2}
\end{figure*}

 % figS3
\begin{figure*}
\begin{center}
\includegraphics[width=\textwidth]{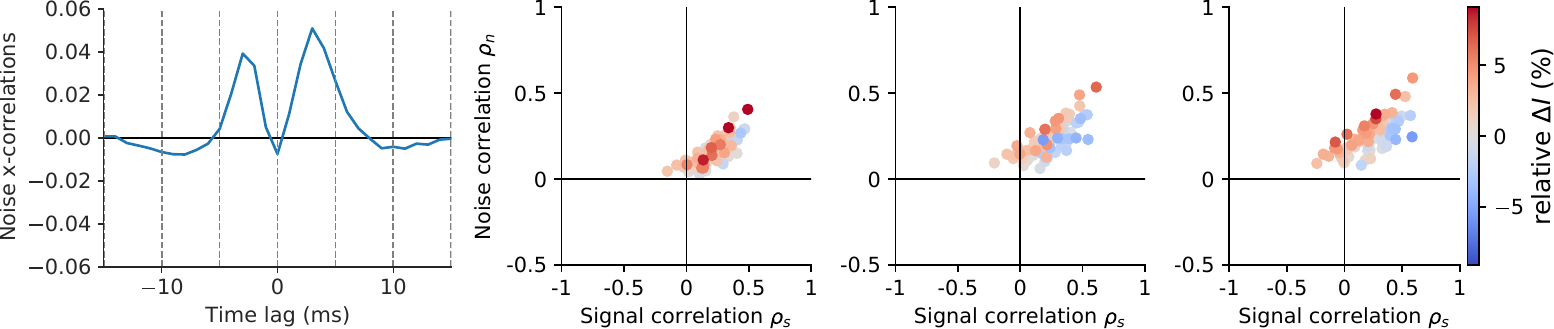}
\end{center}
\caption{\TM
{\bf Choice of bin size.} Left: Example cross-correlogram between two
neighboring retinal ganglion cells. The two symmetric peaks, which are
attributed to the effect of gap junctions, cover a short time scale of
10 ms. Right: Analysis of noise synergy $\Delta I$ (same as Fig. 1F
bottom) for a time bin of 10ms, 20ms, and 30ms, from left to right.
}
\label{fig:figS2}
\end{figure*}

 % figS4
\begin{figure*}
\begin{center}
\includegraphics[width=0.4\textwidth]{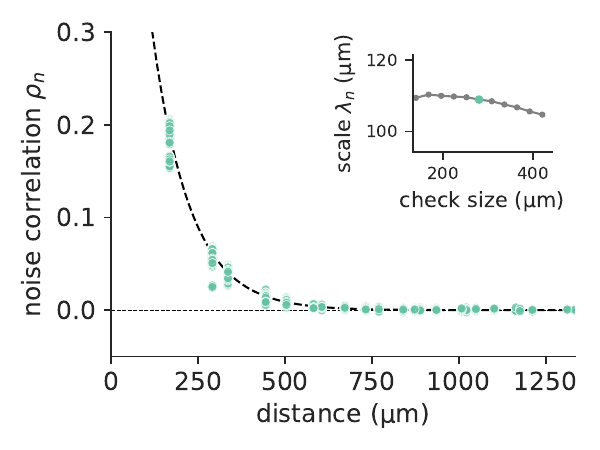}
\end{center}
\caption{\TM
{\bf Correlation structure in augmented data.} Noise correlations in
data generated from the convolutional neural network model (obtained
with time bin of 50 ms). An
exponential fit as a function of distance between neurons returns a
robust correlation length of 110$\mu$m.
}
\label{fig:figS2}
\end{figure*}

\end{document}